\documentclass[floatfix,twocolumn,showpacs,amssymb,amsmath,prd]{revtex4}
\usepackage{graphicx}
\usepackage{dcolumn}
\usepackage{bm}

\begin{document}
\title{CMB polarization map derived from the WMAP 5 year data through Harmonic Internal Linear Combination}
\author{Jaiseung Kim}
\email{jkim@nbi.dk}
\affiliation{Niels Bohr Institute, Blegdamsvej 17, DK-2100 Copenhagen, Denmark}
\author{Pavel Naselsky}
\affiliation{Niels Bohr Institute, Blegdamsvej 17, DK-2100 Copenhagen, Denmark}
\author{Per Rex Christensen}
\affiliation{Niels Bohr Institute, Blegdamsvej 17, DK-2100 Copenhagen, Denmark}

\date{\today}

\begin{abstract}
We have derived whole-sky CMB polarization maps from the WMAP 5 year polarization data, using the Harmonic Internal Linear Combination (HILC) method. Our HILC method incorporates spatial variability of linear weights in a natural way and yields continuous linear weights over the entire sky. To estimate the power spectrum of HILC maps, we have derived a unbiased quadratic estimator, which is similar to the WMAP team's cross power estimator, but in a more convenient form for HILC maps. From our CMB polarization map, we have obtained TE correlation and E mode power spectra without applying any mask. They are similar to the WMAP team's estimation and consistent with the WMAP best-fit $\Lambda$CDM model. Foreground reduction by HILC method is more effective for high resolution and low noise data. Hence, our HILC method will enable effective foreground reduction in polarization data from the Planck surveyor.
\end{abstract}


\maketitle 
 
\section{Introduction}
The Cosmic Microwave Background (CMB) is expected to be linearly polarized by Thompson scattering at the last scattering surface and after re-ionization. 
Most useful information on re-ionization and primordial gravitational wave can be obtained from CMB polarization on large angular scales \citep{Kamionkowski:Flm,Seljak-Zaldarriaga:Polarization,Seljak:signature_gravity_wave,WMAP3:polarization}. 

Foregrounds degrade the attainable accuracy of cosmological information \citep{Tegmark:Forecasts} and Galactic foregrounds are particularly significant on large angular scales \citep{Tegmark:Foreground,Oliveira-Costa:large_polarization2}. Hence, the ability to clean foreground contamination without relying on masking is of the utmost importance for the study of CMB polarization on large angular scales. 

Recently, multi-frequency polarization data for the whole sky has become available from the Wilkinson Microwave Anisotropy Probe (WMAP) \cite{WMAP3:polarization,WMAP5:basic_result}.
For its cosmological analysis, the WMAP team used foreground-reduced maps obtained by a template fitting method. 
Due to heavy foreground contamination within their Galactic mask, the WMAP team's foreground-reduced polarization maps are not suitable as whole-sky CMB polarization maps.
The WMAP team has also produced a low resolution CMB polarization map by Markov Chain Monte Carlo (MCMC) method \citep{WMAP5:foreground}. This MCMC map contains heavy foreground contamination within the Galactic cut \citep{WMAP5:foreground}, making it unsuitable for whole-sky polarization maps.
In contrast, the Internal Linear Combination (ILC) method is one of best blind approaches available in spite of a few drawbacks. As a result, several variants of ILC methods have been developed and implemented to construct a whole-sky CMB temperature map \citep{Eriksen:ILC,Tegmark:CMB_map,WMAP3:temperature,Kim:HILCT,Park:SILC400,needlet}.
Meanwhile, whole-sky CMB polarization maps are quite scarce at the moment. We have derived a whole-sky CMB polarization map from the WMAP 5 year polarization data through the Harmonic Internal Linear Combination (HILC) method, which is a natural extension of our previous efforts to derive a CMB temperature map \citep{Kim:HILCT}.

The outline of this paper is as follows. 
In Sec. \ref{Stokes}, we discuss the Stokes parameters in the context of an all-sky analysis.
In Sec. \ref{multifrequency}, we briefly discuss foreground reduction method with multi-frequency maps. 
In Sec. \ref{spherical_space}, we derive equations and solutions.
In Sec. \ref{WMAP}, we present the results of applying on the WMAP five year polarization data. 
In Sec. \ref{comparison}, we briefly discuss a few other foreground reduction methods and compare them. 
In Section \ref{Discussion}, we summarize our investigation and present our conclusions. 
In Appendix \ref{estimator}, we discuss the unbiased quadratic estimator of power spectrum that we shall be using, which is similar to the WMAP team's cross power estimator, and in more convenient form for an HILC implementation.

\section{STOKES PARAMETERS}
\label{Stokes}
The Stokes parameters describe the state of polarization \citep{Kraus:Radio_Astronomy,Tools_Radio_Astronomy}.
For an all-sky analysis, they are measured in reference to $(\hat e_\theta,\hat e_\phi)$ \citep{Kamionkowski:Flm,Seljak-Zaldarriaga:Polarization}:
\begin{eqnarray*} 
Q&=&\left\langle E_\theta^2-E_\phi^2\right\rangle,\\
U&=&\left\langle 2E_\theta\,E_\phi\right\rangle,
\end{eqnarray*}
where $\left\langle \ldots\right\rangle$ indicates average over a period of the electromagnetic waves, and $\hat e_\theta$ and $\hat e_\phi$ are basis vectors of spherical coordinates. 

All-sky Stokes parameters are decomposed into spin $\pm2$ spherical harmonics \citep{Seljak-Zaldarriaga:Polarization}:
\begin{eqnarray}
Q(\hat {\mathbf n})\pm i U(\hat {\mathbf n})&=&\sum_{l,m} a_{\pm2,lm}\;{}_{\pm2}Y_{lm}(\hat {\mathbf n}).\label{Q_lm+iU_lm} 
\end{eqnarray}
The spin $\pm2$ spherical harmonic coefficients $a_{\pm 2,lm}$ are further decomposed into E and B mode \citep{Seljak-Zaldarriaga:Polarization,Zaldarriaga:Polarization_Exp}:
\begin{eqnarray}
a_{\pm 2,lm}&=&-(a_{E,lm}\pm i \,a_{B,lm}), \label{a_2lm_EB}
\end{eqnarray}
where 
\begin{eqnarray*}
a_{E,lm}=(-1)^m a_{E,l\,-m}^*,\\
a_{B,lm}=(-1)^m a_{B,l\,-m}^*.
 \end{eqnarray*}

\section{Foreground reduction with multi-frequency maps}
\label{multifrequency}
Polarization at a frequency  $\nu_k$ and pixel $\mathbf x$ is as follows:
\begin{eqnarray}
\lefteqn{Q(\mathbf x,\nu_k)\pm i U(\mathbf x,\nu_k)=}\nonumber\\
&&Q_{\mathrm{cmb}}(\mathbf x)\pm i U_{\mathrm{cmb}}(\mathbf x)\,+Q_{\mathrm{fg}}(\mathbf x,\nu_k)\pm i U_{\mathrm{fg}}(\mathbf x,\nu_k)\nonumber\\
&&+Q_{\mathrm{noise}}(\mathbf x,\nu_k)\pm i U_{\mathrm{noise}}(\mathbf x,\nu_k),\label{QU}
\end{eqnarray}
where `fg' denotes the composite foreground signal.
A natural candidate for the estimator of the CMB polarization map is a linear combination of multi-frequency maps, which is as follows: 
\[\sum_{k} {w^k(\mathbf x)}\,\left(Q(\mathbf x,\nu_k)\pm i U(\mathbf x,\nu_k)\right).\]
To keep the CMB unchanged, we choose linear weights such that the sum of linear weights over frequency channels is equal to unity: 
\begin{eqnarray}
\sum_{k} {w^k(\mathbf x)}=1. \label{nomarlization}
\end{eqnarray}
With Eq.~\ref{QU} and \ref{nomarlization}, it is straightforward to show that
\begin{eqnarray}
\lefteqn{\sum_{k} {w^k(\mathbf x)}\;\left(Q(\mathbf x,\nu_k)\pm i U(\mathbf x,\nu_k)\right)=}\nonumber\\
&&Q_{\mathrm{cmb}}(\mathbf x)\pm i U_{\mathrm{cmb}}(\mathbf x)+\zeta(\mathbf x),\label{ilc_pixel}
\end{eqnarray}
where we have neglected noise and defined $\zeta(\mathbf x)$ by
\begin{eqnarray*}
\zeta(\mathbf x)&=&\sum_{k} {w^k(\mathbf x)}\,\left(Q_{\mathrm{fg}}(\mathbf x,\nu_k)\pm i U_{\mathrm{fg}}(\mathbf x,\nu_k)\right).\\
\end{eqnarray*}
The variance of the linear combination map is
\begin{eqnarray}
\sigma^2&=&\left\langle \left|\sum_{k} {w^k(\mathbf x)}\,\left(Q(\mathbf x,\nu_k)\pm i U(\mathbf x,\nu_k)\right)\right|^2 \right\rangle\nonumber\\
&\approx&C^2+2\left\langle \mathrm{Re}\left[ (Q_{\mathrm{cmb}}(\mathbf x)\pm i U_{\mathrm{cmb}}(\mathbf x))\;\zeta(\mathbf x) \right] \right\rangle\nonumber\\
&&+\left\langle |\zeta(\mathbf x)|^2\right\rangle,\label{sigma_ilc}
\end{eqnarray}
where $\langle\ldots \rangle$ denotes average over the whole sky.
The term $C^2$ in Eq.~\ref{sigma_ilc} is the variance of CMB polarization signal and is, therefore, independent of linear weights because of Eq.~\ref{nomarlization}. Since there is no correlation among CMB and foregrounds, we find $\sigma^2\approx C^2+\left\langle |\zeta(\mathbf x)|^2\right\rangle$, and hence the linear combination map of minimum variance contains minimum residual foregrounds. However, the variance minimization on real-world data proceeds in the way to maximize the cancellation between the residual foreground and CMB \citep{WMAP3:temperature,needlet}, since the cross terms $\langle \mathrm{Re}\left[ (Q_{\mathrm{cmb}}(\mathbf x)\pm i U_{\mathrm{cmb}}(\mathbf x))\;\zeta(\mathbf x) \right]\rangle$ completely vanishes only in the limit of infinite number of pixels. The method to reduce the cross term effect is discussed in \citep{Kim:HILCT}.

The frequency spectrum of real-world foregrounds varies with their positions (see \citep{Foreground:Spectra} for a recent treatment), and a spectral index variation as small as $\sim 10\%$ can have a substantial impact on the linear combination of multi-frequency maps \citep{Tegmark:removing_foreground}.
Therefore, the linear weights should possess some spatial variability. 
At the same time, the linear weights of minimum foregrounds are expected to be spatially coherent on small angular scales to a good approximation. 
Hence, we assume the $w^k(\theta,\phi)$ to be some angular functions, which contain spherical harmonics of multipoles up to some cutoff multipole $l_\mathrm{cutoff}$.
The linear combination map formed from multi-frequency maps
\begin{eqnarray}
\lefteqn{Q(\theta,\phi)\pm i U(\theta,\phi)}\nonumber\\
&=&\sum_i w^k(\theta,\phi)\,\left(Q(\theta,\phi,\nu_k)\pm i U(\theta,\phi,\nu_k)\right),\label{wQU}
\end{eqnarray}
can be rewritten in spherical harmonics through the Clebsch-Gordon relation:
\begin{eqnarray}
\lefteqn{a_{\pm2,LM}=}\label{a_LM}\\
&&(-1)^M\sqrt{\frac{2L+1}{4\pi}}\sum_{l m} \sum_{l' m'}
\sqrt{(2l+1)(2l'+1)}\nonumber\\
&&\times \left(\begin{array}{ccc}l&l' &L\\m&m'&-M\end{array}\right)
\left(\begin{array}{ccc}l&l'&L\\0&\pm2&\mp2\end{array}\right)
\sum_{k} w^k_{lm}\,a^k_{\pm2,l' m'}\nonumber,
\end{eqnarray}
where 
\begin{eqnarray}
a_{\pm2,LM}&=&\int {}_{\pm2}Y^*_{LM}(\theta,\phi)\,\left(Q(\theta,\phi)\pm i U(\theta,\phi)\right)\,d\Omega,\nonumber\\
w^k_{l m}&=&\int Y^*_{l m}(\theta,\phi)\,w^k(\theta,\phi)\,d\Omega,\nonumber\\
a^k_{\pm2,l' m'}&=&\int {}_{\pm2}Y^*_{l' m'}(\theta,\phi)\,\left(Q(\theta,\phi,\nu_k)\pm i U(\theta,\phi,\nu_k)\right)\,d\Omega.\nonumber
\end{eqnarray}
The variance of the linear combination map is equivalently given by 
\[\sigma^2= \sum_{LM} |a_{\pm2,LM}|^2,\]
where $L$ is bounded by the triangular inequalities:
\begin{eqnarray}
|l-l'|<=L<=l+l'. \label{triangle}
\end{eqnarray}
 
The constraint $\sum_i w^k(\theta,\phi)=1$ imposed to preserve the CMB signal is expressed in terms of spherical harmonics as follows:
\begin{eqnarray}
\sum_k w^k_{00}&=&\sqrt{4\pi},\label{w00}\\
\sum_k w^k_{lm}&=&0.\;\;\;(l> 0)\label{wlm}
\end{eqnarray}

\section{Determination of linear weights}
\label{spherical_space}
By minimizing the variance, we shall be able to derive equations that lead to the linear weights of minimum foreground.
Since the function $w^k(\theta,\phi)$ is real-valued, $w^k_{lm}$ obeys the reality condition $w^k_{l\,-m}=(-1)^m {w^k_{lm}}^*$. Therefore, we need to determine only $w^k_{lm}$ ($m\ge 0$).
For computational convenience as well as to accommodate the reality condition, we define the real-valued spherical harmonic coefficients $\tilde{w}^k_{lm}$ as $\mathrm{Re}[w^k_{lm}]$, $\mathrm{Im}[w^k_{lm}]$ for $m\ge0$, $m<0$ respectively.
The constraints given by Eq.~\ref{w00} and \ref{wlm} have the following forms for $\tilde{w}^k_{lm}$:
\begin{eqnarray}
\sum_k \tilde{w}^k_{00}&=&\sqrt{4\pi},\label{w00_constraint}\\
\sum_k \tilde{w}^k_{lm}&=&0\;\;\;(l>0)\label{wlm_constraint}.
\end{eqnarray}
The linear weights of the minimum foreground will minimize the variance $\sum_{LM} |a_{\pm2,LM}|^2$ under the constraints Eq.~\ref{w00_constraint} and \ref{wlm_constraint}. The constrained minimization problem is solved conveniently via the Lagrange multiplier method \citep{Eriksen:ILC,Arfken}.  
With the introduction of Lagrange multipliers $\lambda_{lm}$, it can be shown that the variance is minimized
under the constraints Eq.~\ref{w00_constraint} and \ref{wlm_constraint}, when
\begin{eqnarray}
\frac{\partial \sum\limits_{LM} |a_{\pm2,LM}|^2}{\partial\,\tilde{w}^{k'}_{l'm'}}+\lambda_{00}\frac{\partial\left(-\sqrt{4\pi}+\partial\sum\limits_i \tilde {w}^k_{00}\right)}{\partial\,\tilde{w}^{k'}_{l'm'}}\label{variance_minimized}\\
+\sum\limits_{l>0,m}\lambda_{lm}\frac{\partial\sum\limits_i \tilde {w}^k_{lm}}{\partial\,\tilde{w}^{k'}_{l'm'}}&=&0.\nonumber
\end{eqnarray}
By using Eq.~\ref{a_LM}, it can be then shown that Eq.~\ref{variance_minimized} has the following form:
\begin{eqnarray}
\sum_{klm} \left[\alpha^{k'k}_{l'm'lm}\,\tilde{w}^k_{lm}\right]+\lambda_{l'm'}=0,\label{minimum_wlm}
\end{eqnarray}
where $\alpha^{k'k}_{l'm'lm}$ is
\begin{eqnarray}
\alpha^{k'k}_{l'm'lm}=2\mathrm{Re}\left[\sum_{LM}\hat{\gamma}^*_{k'}(l',m',L,M)\,\hat{\gamma}_{k}(l,m,L,M)\right],\label{alpha}
\end{eqnarray}
and  
$\hat{\gamma}_k(l_1,m_1,l_3,m_3)$ is
\begin{eqnarray}
\left\{\begin{array}{r}
\gamma_k(l_1,m_1,l_3,m_3)+(-1)^{m_1}\gamma_k(l_1,-m_1,l_3,m_3)\\
\gamma_k(l_1,m_1,l_3,m_3)\\
i\left[\gamma_k(l_1,-m_1,l_3,m_3)-(-1)^{m_1}\gamma_k(l_1,m_1,l_3,m_3)\right]\end{array}\right.&&\nonumber
\end{eqnarray}
 for $m_1>0$, $m_1=0$ and $m_1<0$ respectively, and
\begin{eqnarray}
\lefteqn{\gamma_{k}(l_1,m_1,l_3,m_3)=}\label{gamma}\\
&&\sum_{l_2 m_2}(-1)^{m_3}\sqrt{\frac{(2l_1+1)(2l_2+1)(2l_3+1)}{4\pi}}\nonumber\\
&&\times \left(\begin{array}{ccc}l_1&l_2&l_3\\m_1&m_2&-m_3\end{array}\right)
\left(\begin{array}{ccc}l_1&l_2&l_3\\0&\pm 2&\mp 2\end{array}\right)
a^{k}_{\pm2,l_2 m_2}.\nonumber
\end{eqnarray}
Thus, the values of the linear weights of the minimum foreground can be found in terms of their Lagrange multipliers $\lambda_{l'm'}$ by solving the system of simultaneous linear equations given by Eq.~\ref{minimum_wlm}.
The values of the Lagrange's multipliers $\lambda_{l'm'}$ can be easily determined by making the solutions of Eq.~\ref{minimum_wlm} satisfy the constraints Eq.~\ref{w00_constraint} and \ref{wlm_constraint}. 
Eq.~\ref{w00_constraint}, \ref{wlm_constraint} and \ref{minimum_wlm} can be conveniently put in matrix form.
For $f$ frequency channels, in matrix notation Eq.~\ref{minimum_wlm} becomes
\begin{eqnarray}
\mathbf A \cdot \mathbf w&=&-\mathbf{\Pi}^{\mathrm{T}}\mathbf{L},\label{AwL}
\end{eqnarray}
where
\begin{eqnarray*}
\mathbf A_{j'j}&=&\alpha^{k'k}_{l'm'lm},\\
\mathbf w_{j}&=&\tilde{w}^k_{lm},\\
\mathbf L_{j''}&=&\lambda_{l'm'},
\end{eqnarray*}
for $j'=f(l'^2+l'+m')+k'$, $j=f(l^2+l+m)+k$, and $j''=l'^2+l'+m'$.
$\mathbf A$ is a $n\times n$ matrix, and $\mathbf w$ and $\mathbf L$ are column vectors of length $n$, and 
length $n/f$ respectively, where $n$ is the total numbers of $\tilde{w}^k_{lm}$. 

In Eq.~\ref{AwL}, $\mathbf w$ is solved in terms of $n/k$ undetermined Langrange multipliers, provided that $\mathbf A$ is invertible:
\begin{eqnarray}
\mathbf w=-\mathbf A^{-1}\mathbf{\Pi}^{\mathrm{T}}\,\mathbf{L}.\label{wAL}
\end{eqnarray}
The constraints given by Eq.~\ref{w00_constraint} and \ref{wlm_constraint} are:
\begin{eqnarray}
\mathbf \Pi\cdot \mathbf w&=&\mathbf {e} \label{constraint}.
\end{eqnarray}
$\mathbf \Pi$ is a $\frac{n}{f}\times n$ matrix, given by
\begin{eqnarray*}
\mathbf \Pi_{ij}=\
&&\left\{\begin{array}{r@{\quad:\quad}l}
1&f(i-1)+1\le j \le f\,i\\
0& \mathrm {otherwise}\end{array}\right.
\end{eqnarray*}
and $\mathbf e$ is a column vector of length $n/f$, given by
\begin{eqnarray*}
\mathbf e_{j}=
&&\left\{\begin{array}{r@{\quad:\quad}l}
\sqrt{4\pi}&j=1\\0&
j>1 \end{array}\right.
\end{eqnarray*}
With Eq.~\ref{constraint} and \ref{wAL}, the undetermined $n/f$ Langrange multipliers are given by:
\begin{eqnarray}
\mathbf L=-(\mathbf{\Pi}\,\mathbf A^{-1}\,\mathbf{\Pi}^{\mathrm{T}})^{-1} \mathbf e,\label{Lagrange}
\end{eqnarray}
Therefore, linear weights in spherical harmonic space is given as follows:
\begin{eqnarray}
\mathbf w=\mathbf A^{-1}\mathbf{\Pi}^{\mathrm{T}}(\mathbf{\Pi}\,\mathbf A^{-1}\,\mathbf{\Pi}^{\mathrm{T}})^{-1} \mathbf e.\label{w_solution}
\end{eqnarray}
It should be noted that Eq.~\ref{w_solution} is not reduced to $\mathbf w=\mathbf{\Pi}^{-1} \mathbf e$, since $\mathbf{\Pi}$ and $\mathbf{\Pi}^\mathrm{T}$ are not square matrices.

\section{foregrounds reduction for E and B mode polarization}
\label{polarization}
The polarization signal for the foregrounds as well as the CMB can be decomposed into E and B mode, where the E and B mode polarization maps at a frequency $\mathbf \nu_k$ are as follows:
\begin{eqnarray*}
Q_E(\mathbf \nu_k,\hat {\mathbf n})\pm i U_E(\mathbf \nu_k,\hat {\mathbf n})&=&-\sum_{lm} a^k_{E,lm}\;{}_{\pm 2}Y_{lm}(\hat {\mathbf n}),\\
Q_B(\mathbf \nu_k,\hat {\mathbf n})\pm i U_B(\mathbf \nu_k,\hat {\mathbf n})&=&\mp i\sum_{lm} \,a^k_{B,lm}\;{}_{\pm 2}Y_{lm}(\hat {\mathbf n}).
\end{eqnarray*}
Foreground sources for E and B mode polarizations can have distinct astrophysical origins.
Therefore, frequency spectra for the foregrounds of E mode polarization might not be identical with those of the B mode polarization. Hence, we shall assume independent linear weights, $w^k_E$ and $w^k_B$, for the E and B mode, and find them respectively through the minimizations of $\sigma^2_{EE}$ and $\sigma^2_{BB}$, where
\begin{eqnarray*}
\sigma^2_{EE}&=&\langle |Q_E(\hat {\mathbf n})\pm i U_E(\hat {\mathbf n})|^2\rangle\\
\sigma^2_{BB}&=&\langle |Q_B(\hat {\mathbf n})\pm i U_B(\hat {\mathbf n})|^2\rangle
\end{eqnarray*}
To determine $w^k_{E,lm}$ and $w^k_{B,lm}$,  we replace $a^{k}_{\pm2,l_2 m_2}$ with $-a^{k}_{E,l_2 m_2}$ 
and $\mp i\,a^{k}_{B,l_2 m_2}$, respectively, in Eq.~\ref{gamma}.
After determing $w^k_E$ and $w^k_B$, we then construct the following maps:
\begin{eqnarray}
\xi(\hat {\mathbf n})&=&\sum_i w^k_E(\hat {\mathbf n})\,(Q_E(\hat {\mathbf n},\mathbf \nu_k)\pm i U_E(\hat {\mathbf n},\mathbf \nu_k)),\label{xi}\\
\beta(\hat {\mathbf n})&=&\sum_i w^k_B(\hat {\mathbf n})\,(Q_B(\hat {\mathbf n},\mathbf \nu_k)\pm i U_B(\hat {\mathbf n},\mathbf \nu_k))\label{beta}.
\end{eqnarray}
Using Eq.~\ref{nomarlization}, we may show  
\begin{eqnarray*}
\xi(\hat {\mathbf n})&=&Q^{\mathrm{cmb}}_E(\hat {\mathbf n})\pm i U^{\mathrm{cmb}}_E(\hat {\mathbf n})\\
&&+\sum_i w^k_E(\hat {\mathbf n})\,(Q^{\mathrm{fg}}_E(\hat {\mathbf n},\mathbf \nu_k)\pm i U^{\mathrm{fg}}_E(\hat {\mathbf n},\mathbf \nu_k))\\
&&+\sum_i w^k_E(\hat {\mathbf n})\,(Q^{\mathrm{noise}}_E(\hat {\mathbf n},\mathbf \nu_k)\pm i U^{\mathrm{noise}}_E(\hat {\mathbf n},\mathbf \nu_k)),\\
\beta(\hat {\mathbf n})&=&Q^{\mathrm{cmb}}_B(\hat {\mathbf n})\pm i U^{\mathrm{cmb}}_B(\hat {\mathbf n})\\
&&+\sum_i w^k_B(\hat {\mathbf n})\,(Q^{\mathrm{fg}}_B(\hat {\mathbf n},\mathbf \nu_k)\pm i U^{\mathrm{fg}}_B(\hat {\mathbf n},\mathbf \nu_k))\\
&&+\sum_i w^k_B(\hat {\mathbf n})\,(Q^{\mathrm{noise}}_B(\hat {\mathbf n},\mathbf \nu_k)\pm i U^{\mathrm{noise}}_B(\hat {\mathbf n},\mathbf \nu_k)).\\
\end{eqnarray*}
Since the linear weights, $w^k_E(\hat {\mathbf n})$ and $w^k_B(\hat {\mathbf n})$, are spatially varying functions, 
$\xi(\hat {\mathbf n})$ contains the B mode polarization of the foregrounds and the noise, while $\beta(\hat {\mathbf n})$ contains E mode polarization of foregrounds and noise.
Therefore, we filter out the B mode polarization from $\xi(\hat {\mathbf n})$ and the E mode polarization from $\beta(\hat {\mathbf n})$.
We then reconstruct the CMB polarization map as follows:
\begin{eqnarray}
Q(\hat {\mathbf n})\pm i U(\hat {\mathbf n})=\tilde\xi(\hat {\mathbf n})+\tilde\beta(\hat {\mathbf n}),\label{QU_cmb}
\end{eqnarray}
where $\tilde\xi(\hat {\mathbf n})$ and $\tilde\beta(\hat {\mathbf n})$ are then filtered 
$\xi(\hat {\mathbf n})$ and $\beta(\hat {\mathbf n})$ respectively.
\section{Application to the WMAP five year data}
\label{WMAP}
\begin{figure}[htb!]
\includegraphics[scale=.27]{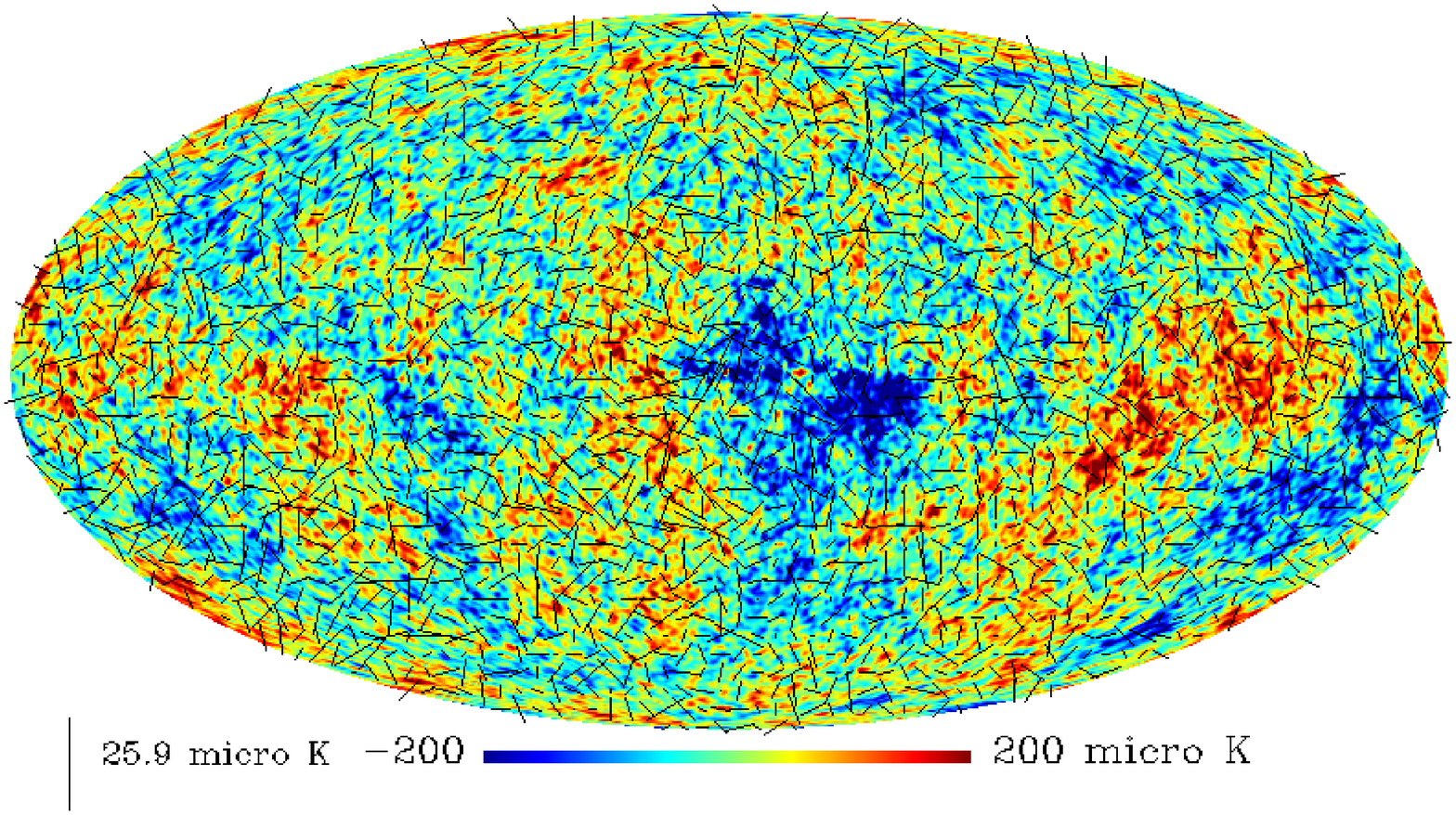}
\includegraphics[scale=.27]{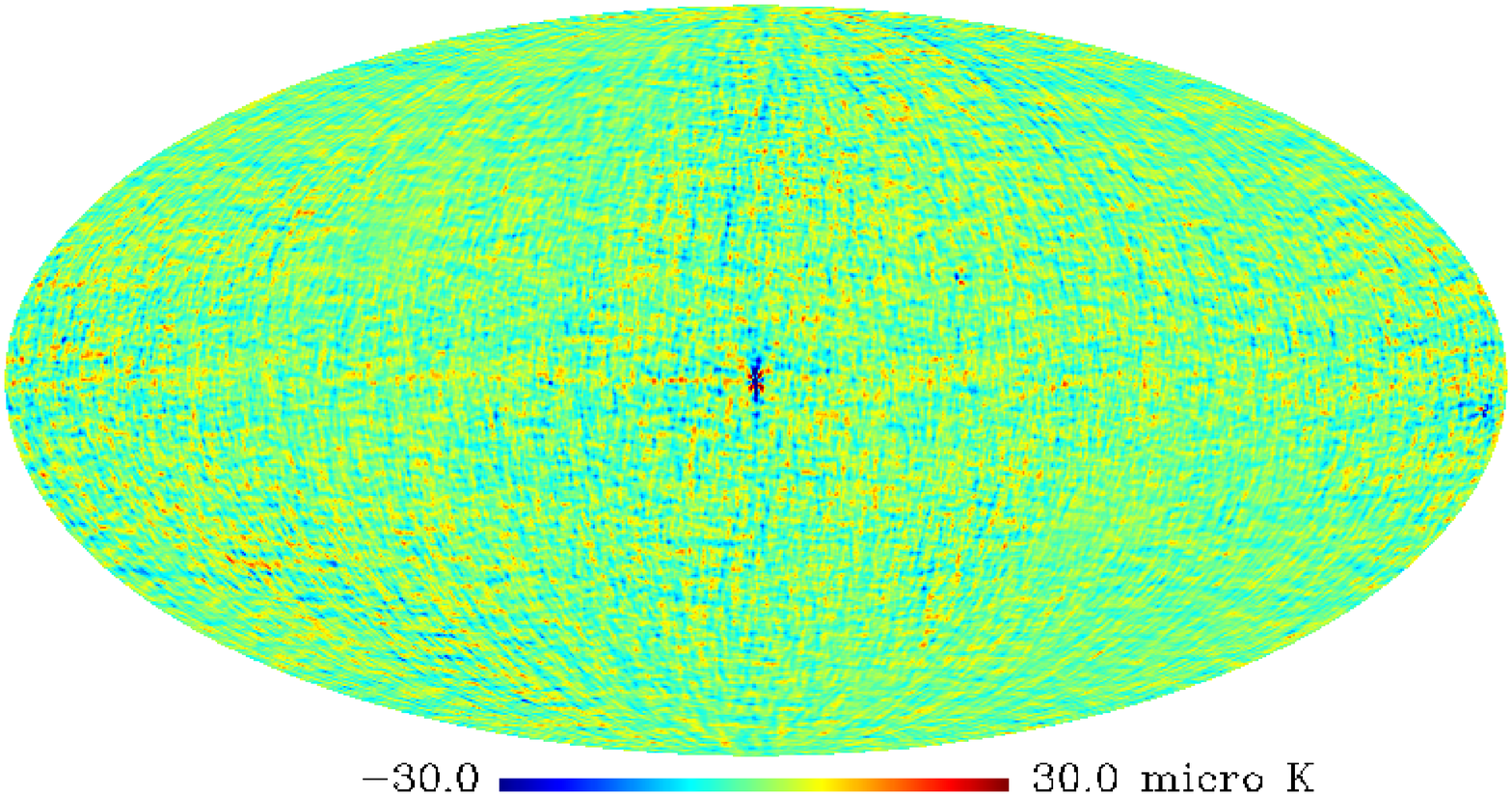}
\includegraphics[scale=.27]{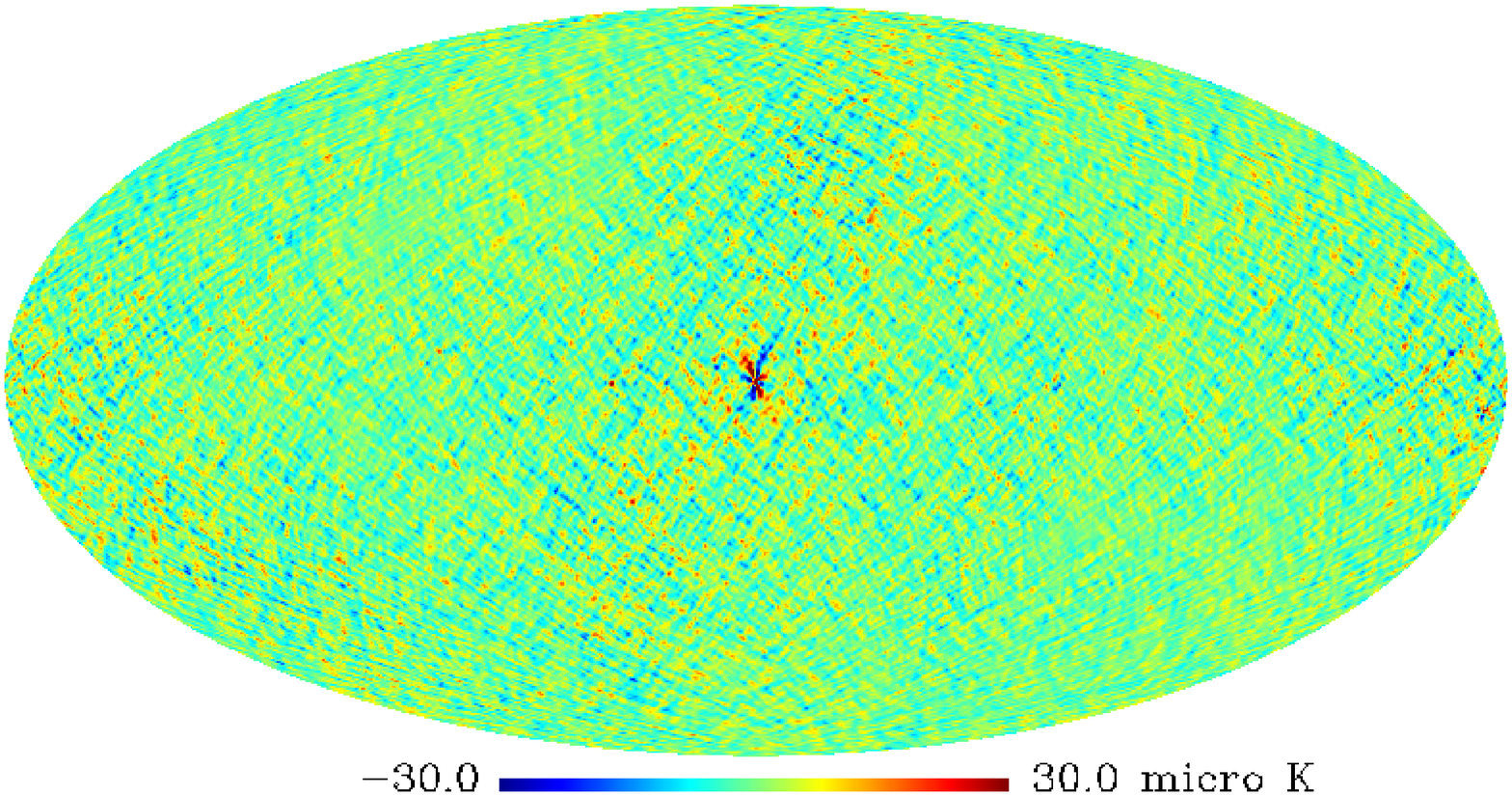}
\includegraphics[scale=.27]{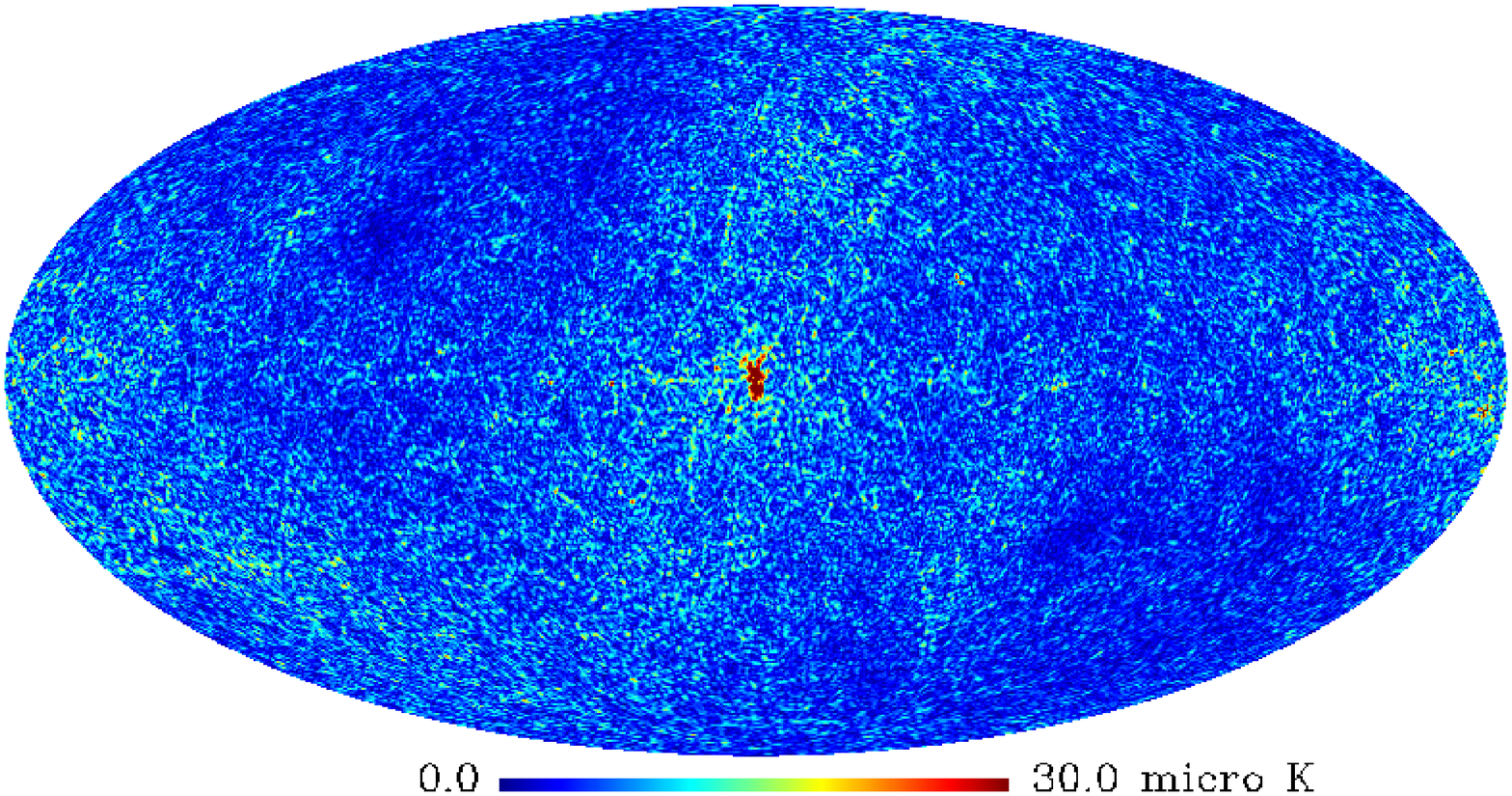}
\caption{the $1^\circ$ FWHM smoothed HILC5YR maps: temperature + polarization, Q, U, and $S=\sqrt{Q^2+U^2}$ (from top to bottom)}
\label{hilc}
\end{figure}
We have applied our foreground reduction method to the WMAP five year polarization data \citep{WMAP5:basic_result}.
Since there exists some anomalous excess power in the W band \citep{WMAP3:polarization,WMAP5:powerspectra}, which is not fully understood, we do not use W band data in the reconstruction of the CMB polarization maps. 
We have considered only the E mode polarization and obtained E mode spherical harmonic coefficients of band maps as follows: \[a^{k}_{E,lm}=\frac{1}{5}\sum_j a^{k,j}_{E,lm}/B^{k}_{l},\]
where $a^{k,j}_{E,lm}$ are the E mode coefficients of the WMAP $j$th year single data 
in the $k$th frequency channel, and the $B^{k}_{l}$ are the beam transfer functions of the WMAP $k$th channel \citep{WMAP5:basic_result,WMAP5:beam,WMAP5:powerspectra}. 

For the frequency channels of the multiple differencing assembly (Q and V band), we have used the average value of the multiple differencing assembly data.
As discussed in Sec. \ref{multifrequency}, the cross terms in Eq.~\ref{sigma_ilc} bias the resulting CMB maps \citep{WMAP3:temperature}. As a part of HILC method, we had developed methods to reduce the cross term effect (see \citep{Kim:HILCT} for details). However, in this work, we do not take special care of the cross term effect, since the cross term effect is negligible, compared to the noise contained in the resulting CMB map.

When linear weights are obtained through variance minimization on noisy data, variance minimization proceeds in the to minimize noise rather than foregrounds. Since the noise of the WMAP data dominates foreground signal on multipoles higher than $\sim 60$, we have used only $a^k_{lm}\;\;(l\le 60)$ in variance minimization (i.e. summation over $l_2$ was done up to $60$ in Eq.~\ref{gamma}). Though ideally the total number of $w^k_{lm}$ may be as high as the total number of available $a^k_{lm}$ (i.e. the number of parameters may be as many as the number of data), we find that the number of $w^k_{lm}$, which keeps the matrices in Eq.~\ref{w_solution} numerically non-singular, is much smaller than the ideal case (i.e. $l_{\mathrm{cutoff}}\ll 60$). This may be attributed to the large bandwidth and the relatively small separation of the WMAP frequency channels, because in such configurations the frequency spectrum of foregrounds may not be numerically distinct enough over the channels.  
Starting with smallest number, we tried various $l_{\mathrm{cutoff}}$ and found the largest and stable $l_{\mathrm{cutoff}}$ for the WMAP polarization data is $5$. Therefore, our linear weights $l_{\mathrm{cutoff}}=5$ is ineffective in cleaning foregrounds, whose frequency spectra vary on the angular scales smaller than $\sim 36^\circ$. 
However, Fig.~\ref{hilc} shows that our CMB polarization maps do not contain significant foregrounds even inside the region corresponding to the WMAP team's Galactic cut. Besides that, we will be able to set $l_{\mathrm{cutoff}}$ as high as the coherence angular scales of the Galactic foreground frequency spectra, when the polarization data from the upcoming Planck surveyor \citep{Planck:sensitivity,Planck:mission} become available.

As a consequence, linear weights of HILC maps are obtained through minimization of $\sigma^2=\sum^{65}_{L=2,M} |a_{\pm2,LM}|^2$ (see Eq.~\ref{triangle}).  However, these linear weights can be applied to band maps containing higher multipoles. Hence, we used $Q_E(\hat {\mathbf n},\mathbf \nu_k)\pm i U_E(\hat {\mathbf n},\mathbf \nu_k )$ of multipoles ($2\le l \le 700$) in the map reconstruction by Eq.~\ref{xi} and \ref{QU_cmb}. 
In Fig.~\ref{hilc}, we show the CMB maps `Harmonic Internal Linear Combination maps' (hereafter, HILC5YR).
The top figure in Fig.~\ref{hilc} shows the CMB polarization as headless vectors whose length is proportional to polarization strength, while the underlying color-coded temperature map is from the HILC5YR temperature map \citep{Kim:HILCT}.
The second and the third figures show the Stokes parameter $Q$ and $U$ respectively, and the bottom figure shows the polarization strength $S=\sqrt{Q^2+U^2}$. All figures shown in Fig.~\ref{hilc} have been generated with $1^\circ$ FWHM beam smoothing to suppress noise for illustrative purpose.
\begin{figure}[htb!]
\centering\includegraphics[scale=.55]{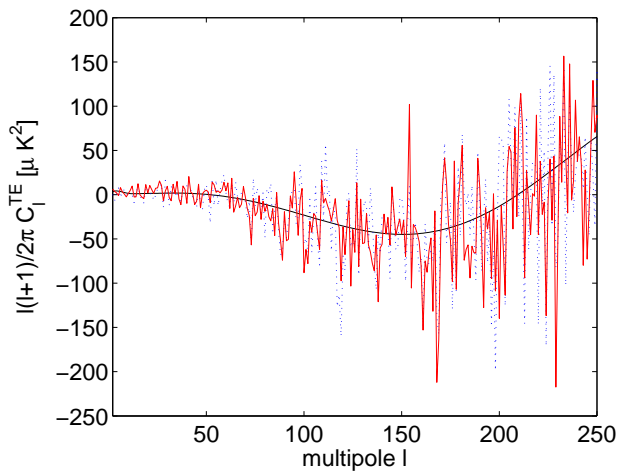}
\caption{TE correlation: solid smooth curve (the WMAP best-fit $\Lambda$CDM model), solid curve (HILC5YR), dotted curve (WMAP team's estimation)}
\label{Cl_TE}
\end{figure}

\begin{figure}[htb!]
\centering\includegraphics[scale=.55]{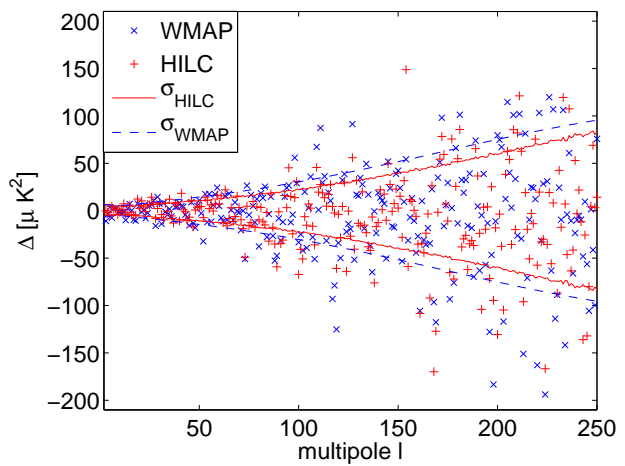}
\caption{$\times$ mark : the difference between the WMAP team's TE correlation and theory, $+$ mark: the difference between HILC TE correlation and theory,
solid curve (the error bars of the HILC TE estimation),
dashed curve (the error bars of the WMAP team's TE estimation)}
\label{Cl_dTE}
\end{figure}

In Fig.~\ref{Cl_TE}, we show the TE correlation of HILC maps and the estimation by the WMAP team.
The theoretical TE correlation of the WMAP best-fit $\Lambda$CDM model is shown as a smooth solid curve in the same figure.
Since noise in a temperature map and in a polarization map is uncorrelated, we have obtained temperature and E mode correlation by a simple quadratic estimator:
\begin{eqnarray}
\hat C^{TE}_l=\frac{1}{2l+1}\sum_m \mathrm{Re} \left[a_{T,lm} {a_{E,lm}}^*\right].\label{Cl_TE_estimator}
\end{eqnarray}
It can be shown that the variance of the estimator $\hat C^{TE}_l$ is
\begin{eqnarray}
\lefteqn{\langle |\hat C^{TE}_l|^2 \rangle}\label{Cl_TE_variance}\\
&=&\frac{1}{2l+1} \left((C^{TE}_l)^2+(C^{TT}_l+N^{TT}_l)(C^{EE}_l+N^{EE}_l)\right),\nonumber
\end{eqnarray}
where $C_l$ and $N_l$ are, respectively, the true power spectra of the underlying CMB and noise distributions.
In Fig.~\ref{Cl_dTE}, we show the difference between the estimations and theory, and estimation errors.
As shown in Fig.~\ref{Cl_dTE}, HILC TE correlations are in agreement with theory and the WMAP team's estimations within expected errors. We have estimated $N^{TT}_l$ and $N^{EE}_l$ from the second term on the right hand side of Eq.~\ref{unbiased_Cl} and obtained the estimation errors by Eq.~\ref{Cl_TE_variance}.

\begin{figure}[htb!]
\centering\includegraphics[scale=.55]{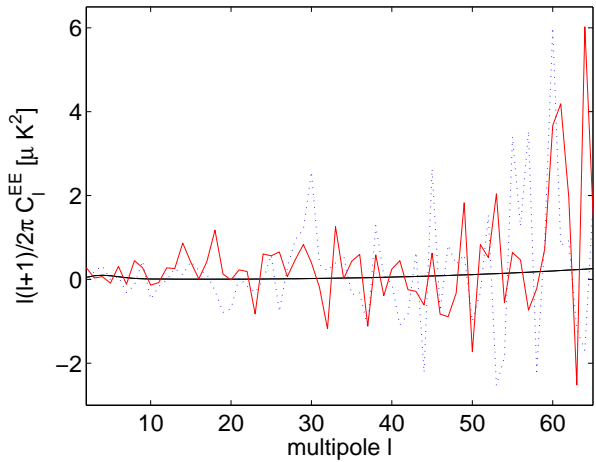}
\caption{E mode power spectrum: solid smooth curve(the WMAP best-fit $\Lambda$CDM model), solid curve(HILC5YR), dotted curve (WMAP team's power spectra)}
\label{Cl_EE}
\end{figure}
\begin{figure}[htb!]
\centering\includegraphics[scale=.55]{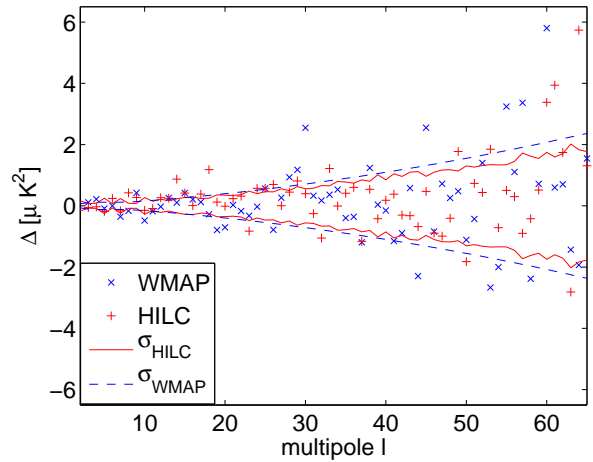}
\caption{$\times$ mark : the difference between the WMAP team's E mode power and theory, $+$ mark: the difference between HILC E mode power and theory,
solid curve (the error bars of the HILC E mode power estimation),
dashed curve (the error bars of the WMAP team's E mode power estimation)}
\label{Cl_dEE}
\end{figure}
On the other hand, we cannot use a simple quadratic estimator for E mode power spectrum because of a noise bias. Hence we have derived a unbiased quadratic estimator of power spectra, which are similar to the WMAP team's cross power spectra, but in a convenient form to use with HILC method. The details on the unbiased quadratic estimator are given in Appendix \ref{estimator}. In Fig.~\ref{Cl_EE}, we show the E mode power spectrum obtained from HILC5YR with our estimator.
We also show the theoretical E mode power spectrum of the WMAP best-fit $\Lambda$CDM model and the WMAP team's estimation \citep{WMAP5:powerspectra} in Fig.~\ref{Cl_EE}.
The large fluctuation in comparison to the theoretical prediction is attributed mostly to the estimation variance associated with noise (i.e. the third term on the right hand side of Eq.~\ref{Cl_variance}). Some estimated E mode power spectra are negative, because they are cross power spectra.
In Fig.~\ref{Cl_dEE}, we show the difference between the estimations and theory, and estimation errors.
We have obtained the estimation errors of HILC E mode power by Eq.~\ref{Cl_variance}. 
As shown in Fig.~\ref{Cl_dEE}, we find that the HILC E mode power estimations are in agreement with theory and the WMAP team's estimations within expected errors. 
\begin{figure}[htb!]
\includegraphics[scale=.2]{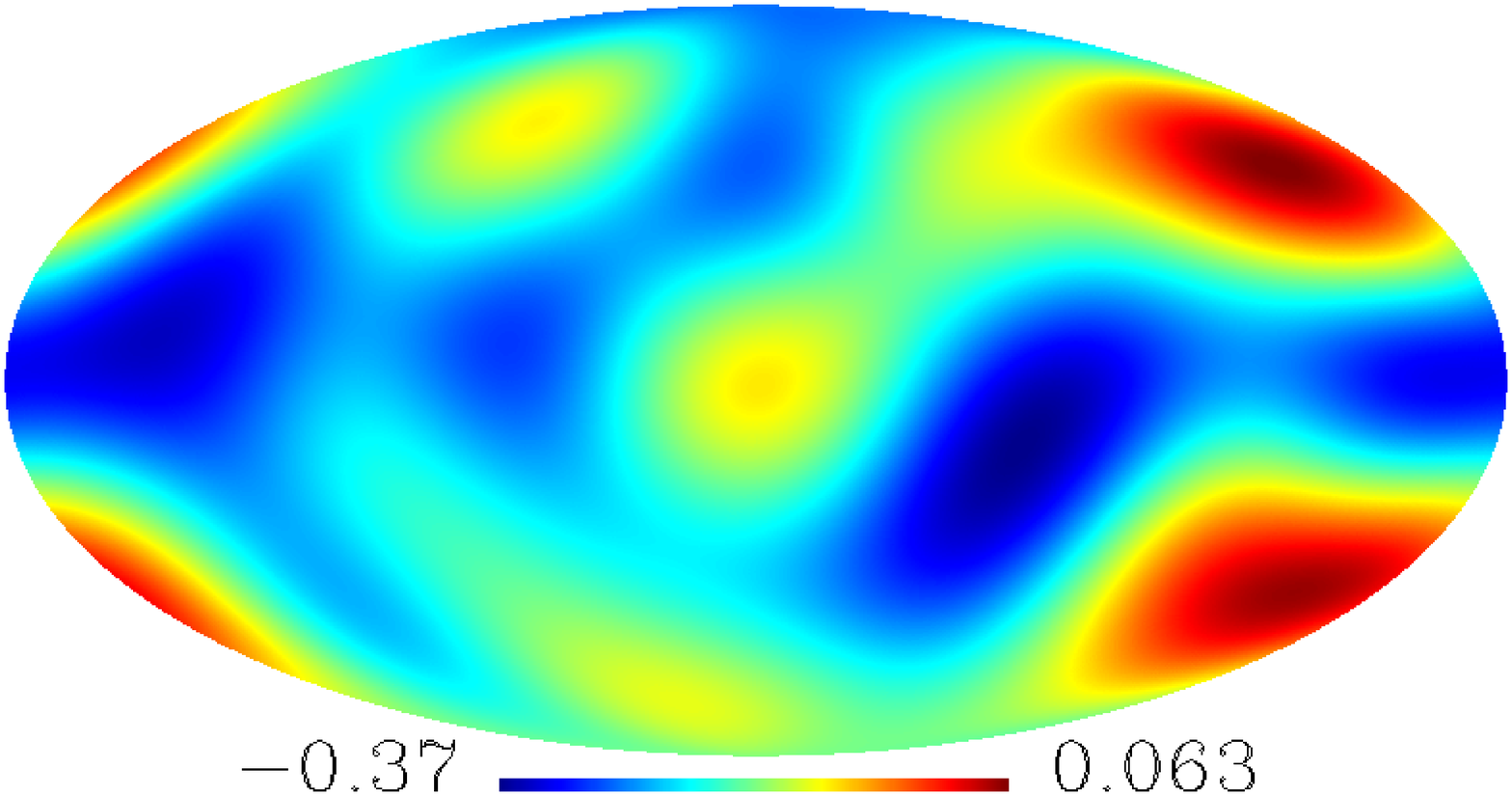}
\includegraphics[scale=.2]{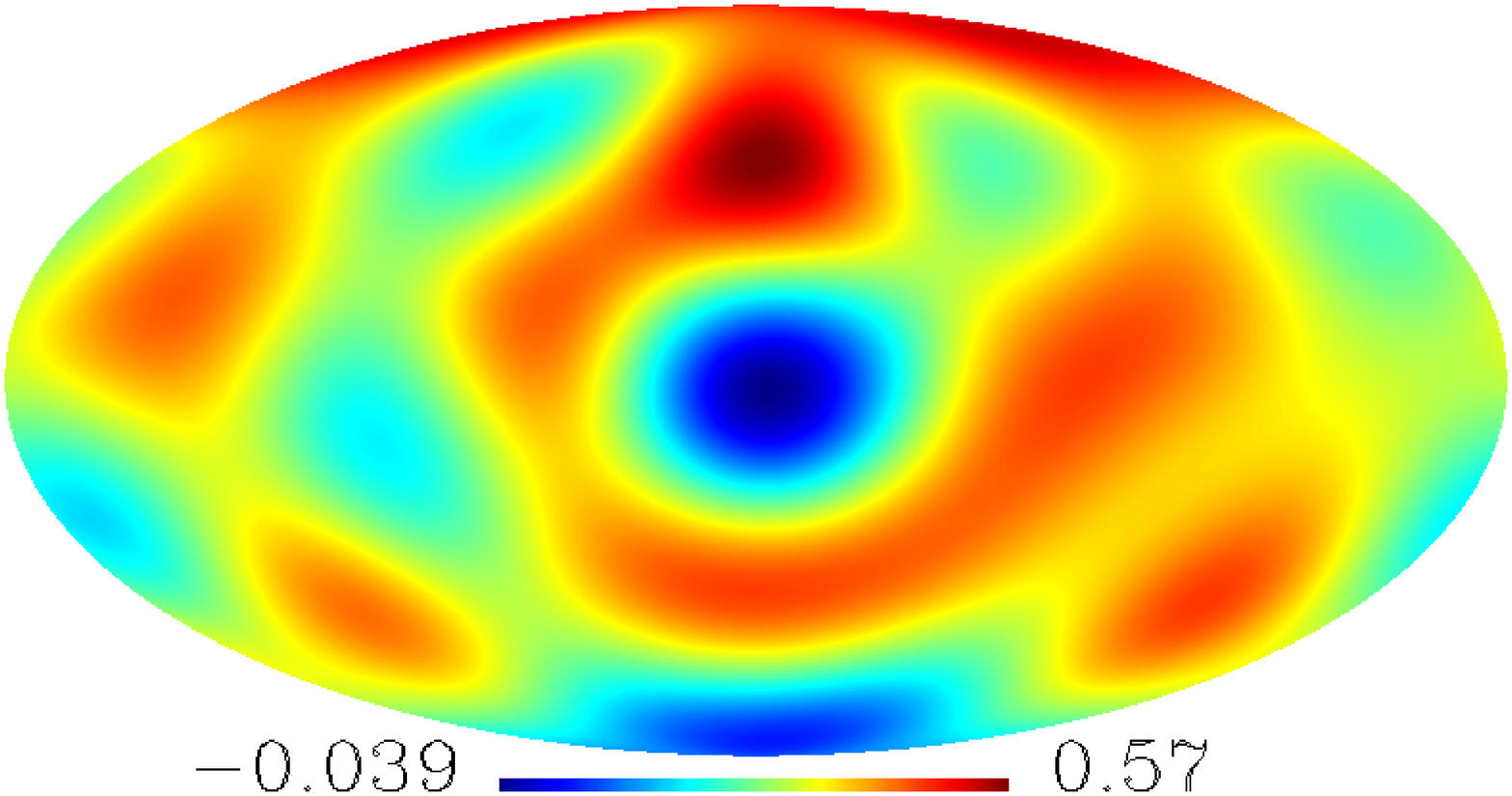}
\includegraphics[scale=.2]{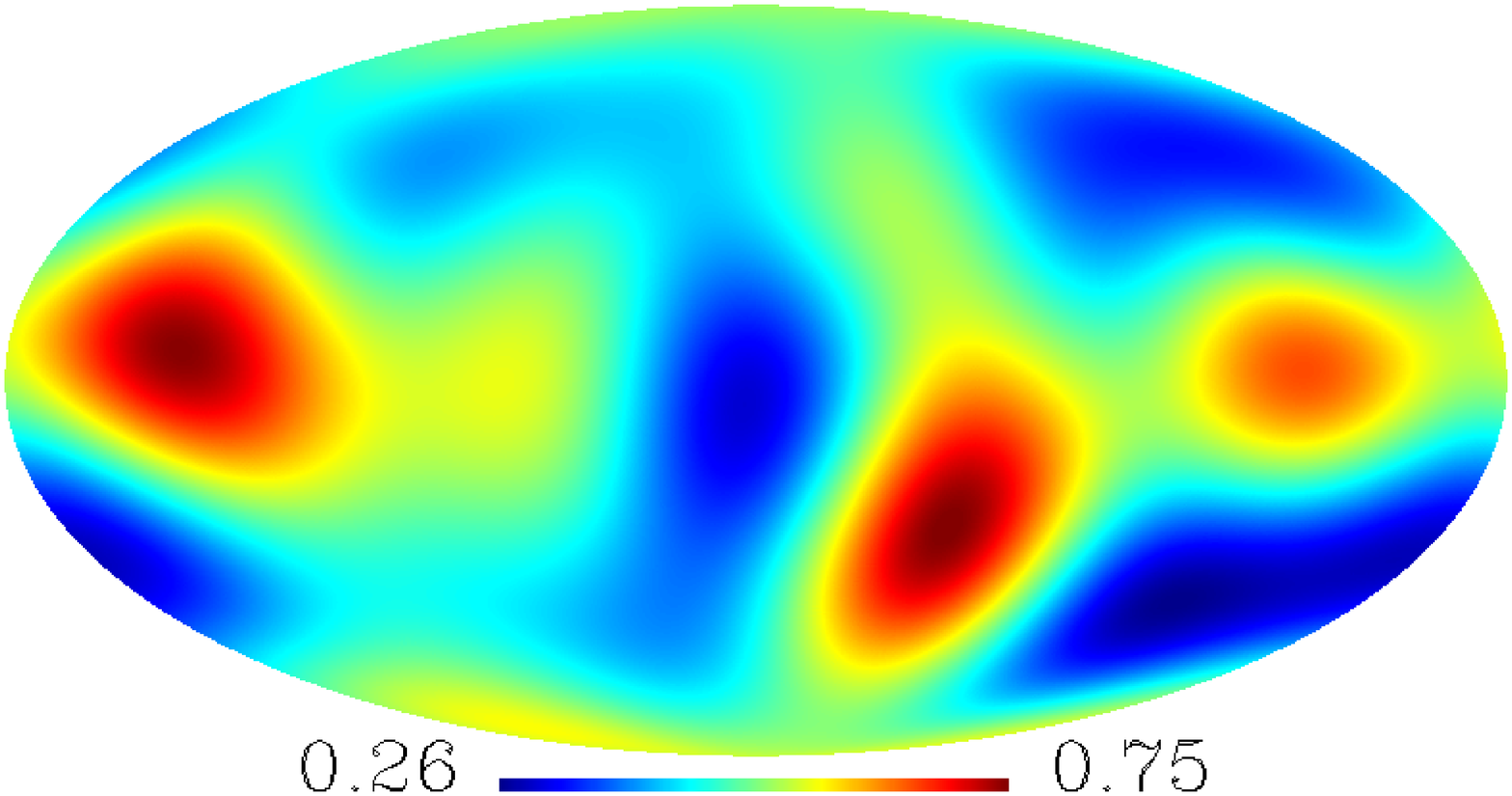}
\includegraphics[scale=.2]{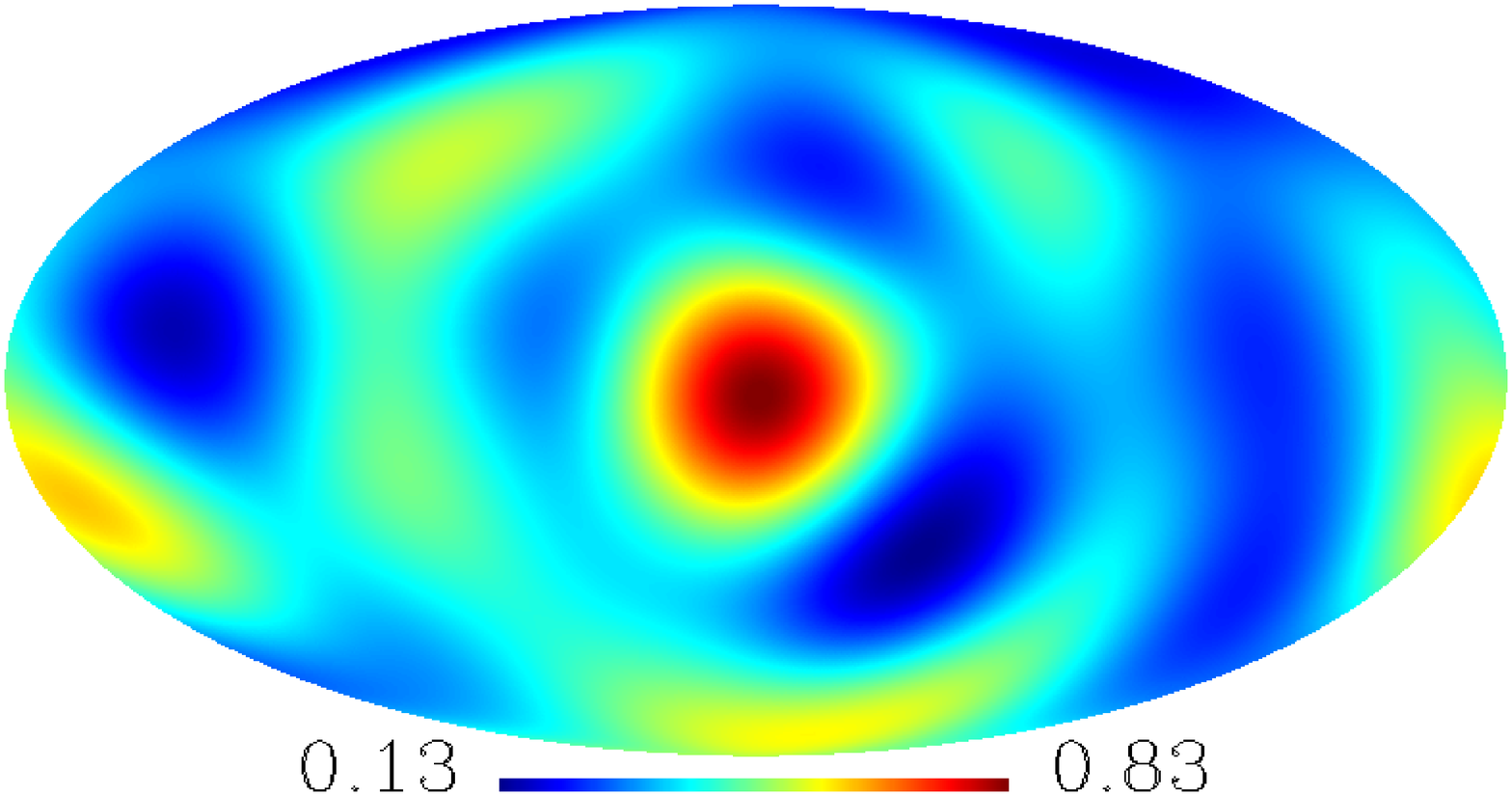}
\caption{The HILC5YR linear weight for K, Ka, Q, and V band (from top to bottom)}
\label{W_HILC}
\end{figure}

The linear weights of HILC5YR, which are continuous over the entire sky, are shown in Fig.~\ref{W_HILC}.
It is worth noting that the linear weights for Q and V bands are positive over the entire sky.
\begin{figure}[htb!]
\centering\includegraphics[scale=.5]{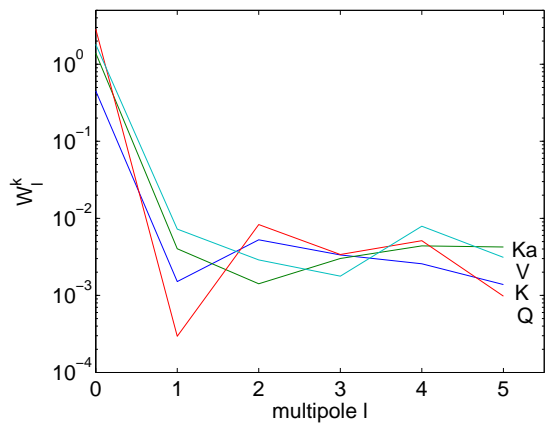}
\caption{Variance of the HILC5YR linear weights}
\label{Wl}
\end{figure}
We have computed the variance of our linear weights by $w^k_l=(2l+1)^{-1}\sum_m |w^k_{lm}|^2$ 
to quantify the spatial variation of our linear weights on different angular scales.
Fig.~\ref{Wl} shows that $w^k_{l}$ tends to decrease with increasing multipole with $w^k_{0}$ being the highest. 

\section{Comparison with other methods}
\label{comparison}
We restrict our discussion to the methods of currently available CMB polarization maps. 
Before proceeding to comparison, we would like to reiterate that the major goal of HILC approach is to reconstruct a whole-sky CMB map, which is important for the study of CMB polarization on large scales. 

The merits of HILC approach are that it naturally incorporates spatial variability of the foreground frequency spectrum in a natural way, and it barely relies on external information about foregrounds. Its major limitation is a poor performance for low Signal-to-Noise Ratio (SNR) data. 

The WMAP team produced foreground reduced maps by a template fitting method, whose model templates are derived from the K band of the WMAP data, and the Finkbeiner dust model 8 \citep{Dust_Extrapolation}. However, there are strong arguments against the interpretation of K band as synchrotron templates \citep{Finkbeiner:Foreground_X,Oliveira-Costa:foreground_X} and it turns out that template models are not sufficient to make good fits simultaneously inside and outside the Galactic cut. Consequently, there is heavy foreground contamination inside the Galactic cut, which makes template-fitted maps unsuitable as a whole CMB sky map.

The WMAP team also produced a low resolution CMB polarization map by the Markov Chain Monte Carlo (MCMC) method \citep{WMAP5:foreground}. The merit of this approach is that it utilizes external information on polarized foreground. 
Its major limitations are that it does not work well within Galactic cut \citep{WMAP5:foreground} and 
its high computational cost makes it unable to produce a high resolution CMB map.

\section{Discussion}
\label{Discussion}
Using the WMAP 5 year polarization data, we have reconstructed whole-sky CMB polarization maps through Harmonic Linear Combination method. From our CMB polarization maps without any masking, we have obtained TE correlation and E mode power spectrum, which are consistent with the WMAP team's estimation. 

Because of the low Signal-to-Noise Ratio of the WMAP polarization data, our HILC polarization maps contain relatively significant noise. Besides that, the low Signal-to-Noise Ratio of the multi-frequency data seriously degrades the foreground reduction of our method, which is relevant to all variants of ILC method. Therefore, we should warn that our CMB polarization maps should be used with some caution.
However, the major aims of this work are to demonstrate HILC on polarization data in preparation for the future Planck data
and to reconstruct whole-sky CMB polarizations maps, whose power spectra are similar to the WMAP team's estimation.

The HILC method is not, in general, less effective in cleaning point sources than in the diffuse Galactic foregrounds, because of assumed finite $l_{\mathrm{cutoff}}$. However, we can effectively suppress the point-source contamination by applying a point-source filter derived by \citep{Tegmark:point_source} at the post-HILC stage. 

In comparison to other foreground cleaning method (e.g. template fitting methods), the effectiveness of our method improves sharply with an increase in Signal-to-Noise Ratio (SNR), the number of frequency channels and the angular resolution of the observation. Therefore, our method will allow us to reduce foregrounds effectively in the polarization data from the upcoming Planck surveyor \citep{Planck:sensitivity,Planck:mission}.

The CMB polarization map, estimated power spectra and linear weights are available from \texttt{http://www.nbi.dk/$\sim$jkim/hilc}.
 
\section{ACKNOWLEDGMENTS}
We are grateful to Hael S. Collins for reading the manuscript and helping us to improve the text.
We are also grateful to the anonymous referee for helpful comments, which led to significant improvements in this paper. 
We acknowledge the use of the Legacy Archive for Microwave Background Data Analysis (LAMBDA). 
Some of the results in this paper have been derived using the HEALPix \citep{HEALPix:Primer,HEALPix:framework} package.
This work is supported by FNU grant 272-06-0417, 272-07-0528 and 21-04-0355. 
\begin{appendix}

\section{unbiased quadratic estimator of power spectra}
\label{estimator}
The estimator derived here is equally applicable to temperature, E mode and B mode power spectra.
Since spherical harmonic coefficients of single-year maps consist of signal and noise terms:
\begin{eqnarray}
a^{k,j}_{lm}=S^{k}_{lm}+N^{k,j}_{lm},\label{alm_ij}
\end{eqnarray}
the spherical harmonic coefficients of band maps, constructed by averaging $n$ single-year maps, are as follows:
\begin{eqnarray*}
a^{k}_{lm}&=&\frac{1}{n}\sum_j a^{k,j}_{E,lm},\\
&=&S^{k}_{lm}+\frac{1}{n}\sum_j N^{k,j}_{lm},
\end{eqnarray*}
where $S^{k}_{lm}$ denote the signal of $k$th frequency channel and $N^{k,j}_{lm}$ denotes the noise of the $j$th year data in the $k$th frequency channel. 
Therefore, it is easy to show that the spherical harmonic coefficient $a_{lm}$ of a HILC map is as follows:
\begin{eqnarray}
a_{lm}=\sum_k w^k_{lm} \otimes S^{k}_{lm}+\frac{1}{n}\sum_{k}\sum_j w^k_{lm} \otimes N^{k,j}_{lm},\label{a_Elm}
\end{eqnarray}
where $\otimes$ denotes spherical harmonic convolution given by Clebsch-Gordon relation.
The spherical harmonic coefficients of a HILC map may be then split into signal and noise terms:
\[a_{lm}=S_{lm}+\frac{1}{n}\sum_j N^{,j}_{lm}.\]
where
$S_{lm}=\sum_k w^k_{lm} \otimes S^{k}_{lm}$ and $N^j_{lm}=\sum_i w^k_{lm} \otimes N^{k,j}_{lm}$.
A simple quadratic estimator for power spectrum gives
\begin{eqnarray}
\lefteqn{(2l+1)^{-1}\sum_m |a_{lm}|^2}\nonumber\\ 
&=&(2l+1)^{-1} \left(\sum_m |S_{lm}|^2 + \frac{1}{n^2}\sum_m \sum_j|N^j_{lm}|^2\right.\nonumber\\
&&+\frac{1}{n}\sum_m \sum_j S_{lm} {N^j}^*_{lm} + {S}^*_{lm}N^j_{lm}\nonumber\\
&&\left.+\frac{1}{n^2}\sum_m \sum_{j>j'} N^j_{lm}{N^{j'}}^*_{lm}+{N^j}^*_{lm} N^{j'}_{lm},\nonumber\right)\\
\label{biased_Cl}
\end{eqnarray}
where the summation is over $1\le j' \le n$ and $j'< j \le n$.
As seen from Eq.~\ref{biased_Cl}, the term $|N^j_{lm}|^2$ is always positive, and hence makes Eq.~\ref{biased_Cl} biased.

Now, consider the following quadratic estimator, which has an extra term:
\begin{eqnarray}
\hat C_l&=&(2l+1)^{-1}\left(\sum_m |a_{lm}|^2\right.\nonumber\\
&&\left.-\frac{1}{n^2(n-1)}\sum_m\sum_{j>j'} |a^{,j}_{lm}-a^{,j'}_{lm}|^2\right)\label{unbiased_Cl}
\end{eqnarray}
where
\begin{eqnarray*}
a^{,j}_{lm}=\sum_k w^k_{lm} \otimes a^{k,j}_{lm},\;\;
a^{,j'}_{lm}=\sum_k w^k_{lm} \otimes a^{k,j'}_{lm}.
\end{eqnarray*}
Using Eq.~\ref{alm_ij}, \ref{biased_Cl} and \ref{unbiased_Cl}, we show that 
\begin{eqnarray}
\hat C_l &=&(2l+1)^{-1} \left(\sum_m |S_{lm}|^2\right.\nonumber\\
&&+\frac{1}{n}\sum_m \sum_j S_{lm} {N^j}^*_{lm} + {S}^*_{lm}N^j_{lm}\nonumber\\
&&\left.+\frac{1}{n(n-1)}\sum_m \sum_{j> j'} N^j_{lm}{N^{j'}}^*_{lm}+{N^j}^*_{lm} N^{j'}_{lm}\nonumber\right)\\
\label{unbiased_Cl2}
\end{eqnarray}
Since the expectation value of $\hat C_l$ is
\[\left\langle \hat C_l \right\rangle =(2l+1)^{-1} \left\langle\sum_m |S_{lm}|^2\right\rangle,\]
we find $\hat C_l$ unbiased.
Now, let us consider the variance of the estimator $\hat C_l$. 
We neglect residual foregrounds in signal $S_{lm}$, assuming foreground reduction is very effective.
Since $\sum_m |S_{lm}|^2/C_l$ follows $\chi^2$ distribution of $2l+1$ degrees of freedom, the variance of the first term on the right hand side of Eq.~\ref{unbiased_Cl2} is
\begin{eqnarray}
\langle |\sum_m |S_{lm}|^2|^2 \rangle=2(2l+1) (C_l)^2. \label{chi2}
\end{eqnarray}
Taking  into account the reality conditions $S_{lm}=(-1)^l S^*_{l\,-m}$ and $N^j_{lm}=(-1)^l {N^j}^*_{l\,-m}$, we find the variances of the second term and the third term on the right hand side of Eq.~\ref{unbiased_Cl2} are respectively:
\begin{eqnarray}
\langle |\sum_m \sum_j S_{lm} {N^j}^*_{lm} + {S}^*_{lm}N^j_{lm}|^2\rangle
=(2l+1)\cdot n\cdot 4\cdot \,C_{l}\,N_{l},\nonumber\\\label{SN}
\end{eqnarray}
and
\begin{eqnarray}
\lefteqn{\langle|\sum_m \sum_{j> j'} N^j_{lm}{N^{j'}}^*_{lm}+{N^j}^*_{lm} N^{j'}_{lm}|^2\rangle}\nonumber\\
&=&(2l+1)\cdot \frac{n(n-1)}{2}\cdot 4\cdot\,(N_{l})^2,\label{NN}
\end{eqnarray}
where $C_l$ and $N_l$ are the theoretical power spectrum of CMB and noise respectively.
In deriving Eq.~\ref{SN} and \ref{NN}, we have made the approximation $\langle{N^j}^*_{lm} N^{j}_{lm'}\rangle \propto \delta_{m m'}$, which is not true for real observations because of anisotropic noise. Hence our estimation on the variance of $\hat C_l$ will be slightly lower than the true values.
Noting there is no correlation among the first, the second, and the third terms on the the right hand side of Eq.~\ref{unbiased_Cl2}, we can show that the estimation variance of $\hat C_l$ is
\begin{eqnarray}
(\Delta \hat C_l)^2 &=& \frac{2}{2l+1}((C_{l})^2 +\frac{2}{n} C_{l} N_{l} + \frac{1}{n(n-1)} (N_{l})^2).\nonumber\\\label{Cl_variance}
\end{eqnarray}

We can also obtain the same result with $\hat C_l$ by applying the WMAP team's cross-power estimator to a pair of distinct single-year maps and averaging the estimation over all combinations. 
However, $\hat C_l$ by Eq.~\ref{unbiased_Cl} is more convenient for an HILC implementation.
\end{appendix}
\bibliographystyle{unsrt}
\bibliography{/home/tac/jkim/Documents/bibliography}
\end{document}